\documentclass{iopart}
\usepackage{iopams}
\usepackage{setstack}
\usepackage{amssymb}
\usepackage{bm}
\usepackage{graphicx}
\begin{document}
\title[Helicity amplitudes, polarization of EM waves and Stokes parameters]{Helicity amplitudes, polarization of EM waves and Stokes parameters: Classical vs. quantum theory}
\author{Iwo Bialynicki-Birula}
\address{Center for Theoretical Physics, Polish Academy of Sciences\\
Aleja Lotnik\'ow 32/46, 02-668 Warsaw, Poland}

\begin{abstract}
It is shown that the helicity amplitudes can be used to describe and analyze the properties of the electromagnetic field in classical and in quantum theory. On the one hand they embody the relativistic content of electromagnetic theory. On the other hand they give a concise description of such experimentally important notions as polarization, the Stokes parameters and the Poincar\'e sphere.
\end{abstract}

\submitto{Journal of Optics}
\vspace{0.5cm}

\section{Introduction}

The purpose of this work is to show that the systematic use of the helicity amplitudes clarifies and unifies the description of the polarization of electromagnetic waves. In our previous publications \cite{bb0,bb1} we have shown how to connect the properties of classical electromagnetic waves with the states of photons described by the helicity amplitudes. In this work, this general approach is applied to the problem of polarization characterized by the Stokes parameters and the Poincar\'e sphere. In particular, the relativistic properties of these parameters are exhibited. The helicity amplitudes are a very convenient tool because they offer a unified description in classical and quantum theory.

The polarization of the electromagnetic waves is usually described in terms of the classical Maxwell theory. However, some of the properties of this description are better seen in quantum electrodynamics. Therefore, both descriptions will be used here and the helicity amplitudes will play a dual role. In the classical theory they just give an alternative description of the electric and magnetic field vectors. In the quantum theory they become the creation and annihilation operators. The whole information about the form of the electromagnetic field is contained in the state vector. There is a new element that appears in the quantum theory: the problem of measurability of various properties (like, for example the Stokes parameters) of the electromagnetic waves. Identifying these properties with the corresponding quantum operators one finds limitations on their simultaneous measurability.

\section{Helicity and Stokes parameters in classical theory}

The starting point is the representation of the electromagnetic field in terms of Riemann-Silberstein (RS) vector ${\bm F}$,
\begin{eqnarray}\label{rs}
{\bm F}(\bm r,t)=\frac{{\bm D}(\bm r,t)}{\sqrt{2\epsilon_0}}+\rmi\frac{{\bm B}(\bm r,t)}{\sqrt{2\mu_0}}.
\end{eqnarray}
All solutions of Maxwell equations in empty space can be expressed \cite{bb0,bb2} in terms of two complex functions $f_\pm(\bm k)$ of the wave vector which appear in the Fourier representation of ${\bm F}$,
\begin{eqnarray}\label{fourier}
{\bm F}(\bm r,t)=\int\!\frac{d^3k}{(2\pi)^{3/2}}{\bm e}(\bm k)\left[f_+(\bm k)\rme^{\rmi\bm k\cdot\bm r-\rmi\omega t}+f_-^*(\bm k)\rme^{-\rmi\bm k\cdot\bm r+\rmi\omega t}\right],
\end{eqnarray}
where ${\bm e}(\bm k)$ is a normalized complex vector that satisfies the equations:
\begin{eqnarray}\label{e}
{\bm k}\times{\bm e}(\bm k)=-\rmi|{\bm k}|{\bm e}(\bm k),\quad {\bm e}^*(\bm k)\cdot{\bm e}(\bm k)=1.
\end{eqnarray}
The polarization vector ${\bm e}(\bm k)$ is defined by these equations up to a phase factor. A convenient choice is:
\begin{eqnarray}\label{ee}
{\bm e}(\bm k)=\frac{1}{\sqrt{2}}\left[\begin{array}{c}
\cos\theta\cos\phi-\rmi\sin\phi\\
\cos\theta\sin\phi+\rmi\cos\phi\\
-\sin\theta
\end{array}\right],
\end{eqnarray}
where $\theta$ and $\phi$ are the directional angles of the wave vector $\bm k$.

The formula (\ref{fourier}) gives a separation of the solution of Maxwell equations into its positive helicity (the first term) and negative helicity (the second term) parts. The helicity amplitudes $f_\pm(\bm k)$ are one-dimensional representations of the 15-parameter conformal group \cite{wig,qed}. These transformations include four translations, three rotations, three special Lorentz transformations, four conformal accelerations, and one dilation.

The 10 generators of the transformations of the Poincar\'e group build from the helicity amplitudes \cite{qed} are at the same time important dynamical quantities,
\numparts
\begin{eqnarray}
\fl\qquad{\rm Energy}\quad H=\sum_\lambda\int\!\frac{d^3k}{k}\,f_\lambda^*({\bm k})\,k\,f_\lambda({\bm k}),\label{genca}\\
\fl\qquad{\rm Momentum}\quad{\bm P}=\frac{1}{c}\sum_\lambda\int\!\frac{d^3k}{k}\,f_\lambda^*({\bm k})\,{\bm k}\,f_\lambda({\bm k}),\label{gencb}\\
\fl\qquad{\rm Angular\,momentum}\quad{\bm M}=\frac{1}{c}\sum_\lambda\int\!\frac{d^3k}{k}\,f_\lambda^*({\bm k})\,\left(\rmi\bm D_\lambda\times{\bm k}+\lambda{\bm k}/|{\bm k}|\right)f_{\lambda}(\bm k),\label{gencc}\\
\fl\qquad{\rm Center\,of\,energy}\quad{\bm N}=\sum_\lambda\int\!\frac{d^3k}{k}\,f_\lambda^*({\bm k})\,k\,i\bm D_\lambda\,f_\lambda(\bm k),\label{gencd}
\end{eqnarray}
\endnumparts
where $\lambda=\pm$. The symbol ${\bm D}_\lambda$ denotes the covariant derivative on the light cone \cite{qed},
\begin{eqnarray}
{\bm D}_\lambda={\bm\partial}-i\lambda{\bm\alpha}({\bm k}),\label{cder}\\
{\bm\alpha}({\bm k})= \sum_{i=1}^3 e_i^*(\bm k){\bm\partial}e_i(\bm k),\label{alpha}
\end{eqnarray}
where ${\bm\partial}$ denotes the nabla in the ${\bm k}$ space. The formulas for the generators are obtained from the standard expressions by inserting the Fourier representation (\ref{fourier}) of the RS vector into the integrals:
\numparts
\begin{eqnarray}
H=\int\!d^3r\,{\bm F}^*({\bm r},t)\!\cdot\!{\bm F}({\bm r},t),\label{gena}\\
{\bm P}=\frac{1}{\rmi c}\int\!d^3r\,{\bm F}^*({\bm r},t)\!\times\!{\bm F}({\bm r},t),\label{genb}\\
{\bm M}=\frac{1}{\rmi c}\int\!d^3r\,{\bm r}\!\times\!\left({\bm F}^*({\bm r},t)\!\times\!{\bm F}({\bm r},t)\right),\label{genc}\\
{\bm N}=\int\!d^3r\,{\bm r}{\bm F}^*({\bm r},t)\!\cdot\!{\bm F}({\bm r},t)-c^2{\bm P}t.\label{gend}
\end{eqnarray}
\endnumparts

The transformation properties of the helicity amplitudes are determined by the transformation properties of the electromagnetic field. They coincide, of course, with the transformation properties that follow from the Wigner theory of representation \cite{wig}. Under the space-time translation by the four-vector $\{t_0,{\bm r}_0\}$ they just pick up the phase,
\begin{eqnarray}\label{trans}
f_\lambda({\bm k})\to \rme^{\rmi\bm k\cdot{\bm r}_0-\rmi\omega t_0}f_\lambda({\bm k}).
\end{eqnarray}
Under the rotation and Lorentz transformations the changes involve the appropriate change of the argument and the multiplication by some factor.

It follows from the transformation (\ref{trans}) of the helicity amplitudes under translations that every bilinear product of the helicity amplitudes is a constant of motion. Energy and momentum clearly fall into this category. In the case of the angular momentum the situation is not so obvious due to presence of the derivative which produces the term proportional to ${\bm k}t_0$. However, this term vanishes upon the evaluation of the vector product.

Looking at the formulas for the generators we notice that there are four even simpler expressions that represent constant of motion of the electromagnetic field,
\numparts
\begin{eqnarray}
S_0=\int\!
\frac{d^3k}{k}\left(f^*_+({\bm k})f_+({\bm k})+f^*_-({\bm k})f_-({\bm k})\right),\label{sima}\\
S_1=\int\!
\frac{d^3k}{k}\left(f^*_+({\bm k})f_-({\bm k})+f^*_-({\bm k})f_+({\bm k})\right),\\
S_2=-\rmi\int\!
\frac{d^3k}{k}\left(f^*_+({\bm k})f_-({\bm k})-f^*_-({\bm k})f_+({\bm k})\right),\\
S_3=\int\!
\frac{d^3k}{k}\left(f^*_+({\bm k})f_+({\bm k})-f^*_-({\bm k})f_-({\bm k})\right).\label{simd}
\end{eqnarray}
\endnumparts
The quantities $S_0$ and $S_3$ appeared before, albeit in a different form. The dimensionless expression $S_0/\hbar c$ written in terms of electric and magnetic field vectors,
\begin{eqnarray}\label{num}
\fl\qquad\frac{S_0}{\hbar c} = \frac{1}{4\pi^2\hbar c}
 \int\!\!d^3r\!\!\int\!\!d^3r'\!\left({\bm D}({\bm
 r}) \frac{1}{\vert{\bm r}-{\bm r}'\vert^2}\!\cdot\!{\bm E}({\bm r}')+{\bm B}({\bm r}) \frac{1}{\vert{\bm r}-{\bm r}'\vert^2}\!\cdot\!{\bm H}({\bm r}') \right),
\end{eqnarray}
has been found long time ago \cite{zeld}) as a measure of photon number. This expression also plays the role of the norm for the photon wave function and it has some remarkable properties. Despite its ``nonrelativistic'' appearance it is invariant not only under the Poincar\'e group but also under the full conformal group \cite{gross}. In turn, the expression for $S_3$ written in terms of electric and magnetic field vectors,
\begin{eqnarray}\label{dual}
\fl\quad\frac{S_3}{\hbar c} = \frac{1}{8\pi\hbar c}\!
 \int\!\!d^3r\!\!\int\!\!d^3r'\!\left(\!{\bm D}({\bm
 r}) \frac{1}{\vert{\bm r}-{\bm r}'\vert}\!\cdot\!{\bm\nabla\!\times\!\bm E}({\bm r}')+{\bm B}({\bm r}) \frac{1}{\vert{\bm r}-{\bm r}'\vert}\!\cdot\!{\bm\nabla\!\times\!\bm H}({\bm r}')\!\right)\!,
\end{eqnarray}
has been found \cite{dt} to be the generator of dual transformations. These properties of $S_0$ and $S_3$ will be confirmed in quantum theory.

One may rewrite the formulas (\ref{sima}-d) in a form reminiscent of the description of spin 1/2 particles by writing down the two helicity in one column \cite{bb0},
\begin{eqnarray}\label{pauli}
\fl\quad S_0=\int\!
\frac{d^3k}{k}{\bm{\mathfrak f}}^\dagger{\bm{\mathfrak f}},\quad
S_1=\int\!
\frac{d^3k}{k}{\bm{\mathfrak f}}^\dagger\sigma_x{\bm{\mathfrak f}},\quad
S_2=\int\!
\frac{d^3k}{k}{\bm{\mathfrak f}}^\dagger\sigma_y{\bm{\mathfrak f}},\quad
S_3=\int\!
\frac{d^3k}{k}{\bm{\mathfrak f}}^\dagger\sigma_z{\bm{\mathfrak f}},\label{col}
\end{eqnarray}
where
\begin{eqnarray}\label{pwf}
{\bm{\mathfrak f}}=\left(\begin{array}{c}f_+({\bm k})\\f_-({\bm k})\end{array}\right),
\end{eqnarray}
and $\{\sigma_x,\sigma_y,\sigma_z\}$ are the Pauli matrices.
The constants of motion (\ref{pauli}) characterize the relative strength of the helicity amplitudes and their phases. The connection with the Stokes parameters can be found from the expression for the electric field vector. The helicity amplitudes $f_\pm(\bm k)$ with the wave vector $\bm k$ generate, according to (\ref{fourier}), the plane wave whose electric field vector is:
\begin{eqnarray}\label{el}
{\bm E}({\bm r},t)=\Re[{\bm e}({\bm k})(f_+\rme^{\rmi\bm k\cdot\bm r-\rmi\omega t}+f^*_-\rme^{-\rmi\bm k\cdot\bm r+\rmi\omega t})].
\end{eqnarray}
In the case of the wave moving in the $z$ direction the complex polarization vector is ${\bm\varepsilon}=\{1,\rmi,0\}/\sqrt{2}$ and the expression (\ref{el}) can be written in the form:
\begin{eqnarray}\label{el1}
\fl\qquad{\bm E}({\bm r},t)=\Re[{\bm\varepsilon}(f_+\rme^{-\rmi(kz-kt)}
+f^*_-\rme^{-\rmi(kz-kt)})]=\Re\left[(f_+{\bm\varepsilon}
+f_-{\bm\varepsilon}^*)\rme^{-\rmi(kz-kt)}\right].
\end{eqnarray}
Comparing this formula with the textbook definition \cite{jack} of the standard Stokes parameters $\{s_0,s_1,s_2,s_3\}$ in the circular polarization basis, we obtain
\begin{eqnarray}\label{stokes}
f_+=|f_+|e^{\rmi\delta_+},\;f_-=|f_-| e^{\rmi\delta_-},\nonumber\\
s_0=|f_+|^2+|f_-|^2,\;s_1=2|f_+||f_-|\cos(\delta_--\delta_+),\nonumber\\
s_2=2|f_+||f_-|\sin(\delta_--\delta_+),\;s_3=|f_+|^2-|f_-|^2.
\end{eqnarray}
These four parameters are not independent since
\begin{eqnarray}\label{rel}
s_0^2=s_1^2+s_2^2+s_3^2.
\end{eqnarray}

As a visual representation of the Stokes parameters one introduces \cite{jack} the ellipses traced by the electric and magnetic field vectors. For every plane wave the ellipses representing the electric and magnetic field vectors differ only by the $90\deg$ rotation. The two numbers $|f_+|$ and $|f_-|$ define the major semi-axis and minor semi-axis of the polarization ellipse while the phase difference $\delta_--\delta_+$ defines the orientation of the ellipse.

In a similar fashion, one may also assign an ellipse to the global Stokes parameters (\ref{sima}-d) for any integrable solution of the Maxwell equations. To this end, the major semi-axis $a$, the minor semi-axis $b$, and the phase difference are defined as follows:
\begin{eqnarray}\label{glob}
a=\sqrt{\frac{S_0+S_3}{2}},\quad b=\sqrt{\frac{S_0-S_3}{2}},\quad e^{i(\delta_--\delta_+)}=\frac{S_1+iS_2}{\sqrt{S_0^2-S_3^2}}.
\end{eqnarray}
In this way, one obtains a single ellipse characterizing the global properties of the electromagnetic field.
In the limit, when the amplitudes $f_\pm(\bm k)$ are concentrated around some wave vector $\bm k$, these general formulas coincide with the expressions (\ref{stokes}) for the plane monochromatic wave.

For monochromatic waves it is also possible to assign ellipses representing the motion of the electric and magnetic field vectors at each point. To this end one may start from the formula (\ref{fourier}) which now takes the form:
\begin{eqnarray}\label{mono}
{\bm F}_\omega(\bm r,t)=\rme^{-\rmi\omega t}{\bm F}_+(\bm r)+\rme^{\rmi\omega t}{\bm F}_-(\bm r),
\end{eqnarray}
and the formulas for the electric and magnetic field vectors read (disregarding some constant coefficients):
\begin{eqnarray}\label{gen}
\fl\qquad{\bm E}(\bm r,t)=\Re[{\bm F}_+(\bm r)+{\bm F}_-(\bm r)]\cos(\omega t)+\Im[{\bm F}_+(\bm r)-{\bm F}_-(\bm r)]\sin(\omega t),\\
\fl\qquad{\bm B}(\bm r,t)=\Im[{\bm F}_+(\bm r)+{\bm F}_-(\bm r)]\cos(\omega t)-\Re[{\bm F}_+(\bm r)-{\bm F}_-(\bm r)]\sin(\omega t).
\end{eqnarray}
In general, the polarization ellipses change from point to point. The ellipses at different points are different and unlike in the case of a plane wave, the ellipse traced by the electric field vector is different from the ellipse traced by the magnetic field vector. The translation of the parameters of the ellipses into the Stokes parameters is still possible with the use of (\ref{stokes}) but the electric and magnetic Stokes parameters will be different and the relations (\ref{rel}) will not hold. Of course, one may define a single set of Stokes parameters for monochromatic waves by restricting the integration in the definition of global Stokes parameters (\ref{sima}-d).

\section{Helicity and Stokes parameters in quantum theory}

The negative helicity amplitude in (\ref{fourier}) was denoted by the complex conjugated function $f_-^*(\bm k)$ for two reasons, both related to the description of the electromagnetic field in quantum theory. In the framework of the first quantization the helicity amplitude $f_-(\bm k)$ (and not $f_-^*(\bm k)$) plays the role of the photon wave function because its time dependence (\ref{trans}) is that for a positive energy particle. This assignment is fully confirmed in the second quantization when the amplitudes $f_\pm(\bm k)$ become photon annihilation operators while $f_\pm^*(\bm k)$ become photon creation operators \cite{bb2,qed}.
\begin{eqnarray}\label{sq}
f_\pm(\bm k)\rightarrow \sqrt{\hbar c}\,a_\pm(\bm k),\quad f^*_\pm(\bm k)\rightarrow \sqrt{\hbar c}\,a^\dagger_\pm(\bm k).
\end{eqnarray}
The multipliers $\sqrt{\hbar c}$ are needed to account for different dimensions of the helicity amplitudes and the operators since the helicity amplitudes carry the dimensionality of the electromagnetic field while the dimensionality of the operators is fixed by their commutation relations,
\begin{eqnarray}\label{com}
[a_\lambda(\bm k), a^\dagger_{\lambda'}(\bm k')]=k\,\delta_{\lambda\lambda'}\delta^{(3)}(\bm k-\bm k').
\end{eqnarray}
The prefactor $k$ is needed here if one wants to keep the same transformation properties for $a_\pm$ as for $f_\pm$.

The substitution (\ref{sq}) produces from the classical RS vector the field operator
\begin{eqnarray}\label{fourierq}
\fl\qquad{\bm\hat{F}}(\bm r,t)=\sqrt{\hbar c}\int\!\frac{d^3k}{(2\pi)^{3/2}}{\bm e}(\bm k)\left[a_+(\bm k)\rme^{\rmi\bm k\cdot\bm r-\rmi\omega t}+a_-^\dagger(\bm k)\rme^{-\rmi\bm k\cdot\bm r+\rmi\omega t}\right].
\end{eqnarray}
The basic physical operators expressed in terms of creation and annihilation operators are:
\numparts
\begin{eqnarray}
\fl{\rm Energy\,operator}\quad \hat{H}=\sum_\lambda\int\!\frac{d^3k}{k}\,a^\dagger_\lambda({\bm k})\,\hbar\omega\,a_\lambda({\bm k}),\label{genqa}\\
\fl{\rm Momentum\,operator}\quad{\bm\hat{ P}}=\sum_\lambda\int\!\frac{d^3k}{k}\,a^\dagger_\lambda({\bm k})\,\hbar{\bm k}\,a_\lambda({\bm k}),\label{genqb}\\
\fl{\rm Angular\,momentum\,operator}\!\quad{\bm\hat{ M}}=\sum_\lambda\int\!\frac{d^3k}{k}\,a^\dagger_\lambda({\bm k})\left(\rmi\hbar\bm D_\lambda\times{\bm k}+\lambda{\bm k}/|{\bm k}|\right)a_\lambda(\bm k),\label{genqc}\\
\fl{\rm Center\,of\,energy\,operator}\quad{\bm\hat{ N}}=\sum_\lambda\int\!\frac{d^3k}{k}\,a^\dagger_\lambda({\bm k})\,\omega\,i\hbar\bm D_\lambda\,a_\lambda(\bm k).\label{genqd}
\end{eqnarray}
\endnumparts

There are two ways to introduce Stokes parameters in quantum theory. One associates these parameters with single photon states while the other treats the whole electromagnetic field as one quantum system.

The Stokes parameters associated with photon states are basically the same as those characterizing the classical electromagnetic waves. In both cases the carriers of the information are the helicity amplitudes. In the classical case, the helicity amplitudes inserted into the formula (\ref{fourier}) define the electromagnetic wave. In the quantum case, the helicity amplitudes define the state of the photon through the formula:
\begin{eqnarray}\label{oneph}
|1_{\rm ph}\rangle=a^\dagger[{\bm{\mathfrak f}}]|0\rangle=\int\!
\frac{d^3k}{k}\left(f_+(\bm k)a^\dagger_+(\bm k)+f_-(\bm k)a^\dagger_-(\bm k)\right)|0\rangle.
\end{eqnarray}
In principle, the photon wave functions appearing in this formula should be normalized to allow for the probabilistic interpretation,
\begin{eqnarray}\label{norm}
\sum_\lambda\int\!
\frac{d^3k}{k}|f_\lambda(\bm k)|^2=1.
\end{eqnarray}
However, in practice one may keep the wave function normalized while restricting it approximately to a single wave vector. Then one may characterize the photon state in terms of the Stokes parameters given by the formulas (\ref{stokes}) as if one was dealing with the plane wave. Similarly, one may extend this description to all (quasi) monochromatic photon states as was done for classical waves.

The combination of helicity amplitudes and the creation operators in (\ref{oneph}) may be used also in the definition of coherent quantum states of electromagnetic radiation. Namely, we may construct the Glauber displacement operator \cite{rg} and define the coherent state as follows:
\begin{eqnarray}\label{coh}
|\rm{coh}\rangle=\exp\left(\sqrt{N}(a^\dagger[{\bm{\mathfrak f}}]-a[{\bm{\mathfrak f}}])\right)|0\rangle,
\end{eqnarray}
where $N$ is the average number of photons in the coherent state. The average value of the electromagnetic field in the coherent state $\langle\rm{coh}|\bm{F}|\rm{coh}\rangle$ is given by (\ref{fourier}) with the extra factor $\sqrt{N}$ in front of the integral.

The direct use of the creation and annihilation operators
brings out novel quantum aspects of helicity and polarization. For a two-mode model the Stokes operators were introduced long time ago \cite{jr} as follows:
\begin{eqnarray}\label{qpauli}
\hat{s}_0={\bm{\mathfrak a}}^\dagger{\bm{\mathfrak a}},\;\hat{s}_1={\bm{\mathfrak a}}^\dagger\sigma_x{\bm{\mathfrak a}},\;
\hat{s}_2={\bm{\mathfrak a}}^\dagger\sigma_y{\bm{\mathfrak a}},\;\hat{s}_3={\bm{\mathfrak a}}^\dagger\sigma_z{\bm{\mathfrak a}},
\end{eqnarray}
where ${\bm{\mathfrak a}^\dagger}=\{a^\dagger_+,a^\dagger_-\}$ and ${\bm{\mathfrak a}}=\{a_+,a_-\}$ are two-component creation and annihilation operators of photons with positive and negative helicity. These quantum Stokes operators satisfy the commutation relations of angular momentum as noted in \cite{jr}.

The commutation relations impose limitations on the measurement of the properties described by these operators. Only one component of the Stokes operator (\ref{qpauli}) may have a well-defined value. Let us consider the component of the Stokes operator ${\bm m}\!\cdot\!\hat{\bm s}$ along the unit vector $\{m_x,m_y,m_z\}$. The states of the photon which belong to the eigenvalues $\pm 1$ of this operator are created by the following combination of creation operators:
\begin{eqnarray}\label{ph}
\fl\qquad|1_{\rm ph},\pm\rangle=\frac{1}{\sqrt{2(1\mp m_z)}}\left(\pm(m_x+\rmi m_y)a^\dagger_++(1\mp m_z)a^\dagger_-\right)|0\rangle.
\end{eqnarray}
The helicity amplitudes (i.e. the photon wave functions in (\ref{oneph})) in this case are:
\begin{eqnarray}\label{ham}
f_+=\frac{\pm(m_x+\rmi m_y)}{\sqrt{2(1\mp m_z)}},\quad f_-=\frac{1\mp m_z}{\sqrt{2(1\mp m_z)}}.
\end{eqnarray}
By comparing this formula with (\ref{stokes}), one finds the relation between the unit vector $\bm m$ and Stokes parameters (\ref{stokes}). Thus, the connection known in classical optics of the Stokes parameters with unit vectors on the Poincar\'e sphere \cite{bw} acquires a new interpretation in terms of the eigenstates (\ref{ph}) of angular momentum operators.

The operators representing the global Stokes parameters (\ref{sima}-d) are:
\numparts
\begin{eqnarray}
\hat{S}_0=\int\!
\frac{d^3k}{k}\left(a^\dagger_+({\bm k})a_+({\bm k})+a^\dagger_-({\bm k})a_-({\bm k})\right),\label{sqa}\\
\hat{S}_1=\int\!
\frac{d^3k}{k}\left(a^\dagger_+({\bm k})a_-({\bm k})+a^\dagger_-({\bm k})a_+({\bm k})\right),\\
\hat{S}_2=-\rmi\int\!
\frac{d^3k}{k}\left(a^\dagger_+({\bm k})a_-({\bm k})-a^\dagger_-({\bm k})a_+({\bm k})\right),\\
\hat{S}_3=\int\!
\frac{d^3k}{k}\left(a^\dagger_+({\bm k})a_+({\bm k})-a^\dagger_-({\bm k})a_-({\bm k})\right),\label{sqd}
\end{eqnarray}
\endnumparts
where we dropped the factor $\hbar c$. The operators $\hat{S}_0$ and $\hat{S}_3$ are of special importance. The operator $\hat{S}_0=\hat{N}_++\hat{N}_-$, acting on a state with $n_+$ right-handed photons and $n_-$ left-handed photons, gives the total number of photons $n=n_++n_-$. This confirms the identification of $S_0$ in (\ref{num}) as the photon number. The operator $\hat{S}_3=\hat{N}_+-\hat{N}_-$ acting on the same state gives, what may be called, the total helicity $n=n_+-n_-$. Total helicity is, at the same time, the generator of duality transformations (\ref{dual}), as shown by the formula:
\begin{eqnarray}\label{dualq}
\rme^{-\rmi\alpha\hat{S}_3}\hat{\bm F}\rme^{\rmi\alpha\hat{S}_3}=\rme^{\rmi\alpha}\hat{\bm F}.
\end{eqnarray}
The multiplication of the RS vector by the phase factor results in the dual rotation of the electric and magnetic field.

In summary, I have shown that the use of helicity amplitudes and the corresponding creation and annihilation operators allows for making a direct connection between the classical and the quantum description of electromagnetic phenomena. In this way, we may establish quantum counterparts of classical properties and vice versa.

\section*{Acknowledgments}

I am grateful to Zofia Bialynicka-Birula for her harsh but constructive and very helpful criticism.

\end{document}